\documentclass[superscriptaddress,twocolumn,aps,floatfix,nopacs]{revtex4}
\usepackage{amsmath,amssymb,eucal,graphicx}
\pdfoutput=1 

\begin{document}

\title{Accumulation of driver and passenger mutations during tumor progression}

\author{Ivana Bozic}
\author{Tibor Antal}
\affiliation{Program for Evolutionary Dynamics, Department of Mathematics, Department of Organismic and Evolutionary Biology, Harvard University, Cambridge, Massachusetts 02138, USA.}
\author{Hisashi Ohtsuki}
\affiliation{Department of Value and Decision Science, Tokio Institute of Technology, Tokio 152-8552, Japan.}
\author{Hannah Carter}
\author{Dewey Kim}
\affiliation{Department of Biomedical Engineering, Institute for Computational Medicine, Johns Hopkins University, Baltimore, Maryland 21218, USA.}
\author{Sining Chen}
\affiliation{Department of Biostatistics, School of Public Health, University of Medicine and Dentistry of New Jersey, Piscataway, New Jersey 08854, USA.}
\author{Rachel Karchin}
\affiliation{Department of Biomedical Engineering, Institute for Computational Medicine, Johns Hopkins University, Baltimore, Maryland 21218, USA.}
\author{Kenneth W. Kinzler}
\author{Bert Vogelstein}
\affiliation{Ludwig Center for Cancer Genetics and Therapeutics, and Howard Hudges Medical Institute at Johns Hopkins Kimmel Cancer Center, Baltimore, Maryland 21231, USA.}
\author{Martin A. Nowak}
\affiliation{Program for Evolutionary Dynamics, Department of Mathematics, Department of Organismic and Evolutionary Biology, Harvard University, Cambridge, Massachusetts 02138, USA.}


%

\begin{abstract}
Major efforts to sequence cancer genomes are now occurring throughout the world\cite{simpson09,collins07}. Though the  emerging data from these studies are illuminating, their reconciliation with epidemiologic and clinical observations poses a major challenge\cite{teschendorff09}.  In the current study, we provide a novel mathematical model that begins to address this challenge. We model tumors as a discrete time branching process\cite{athreya72} that starts with a single driver mutation and proceeds as each new driver mutation leads to a slightly increased rate of clonal expansion.   Using the model, we observe tremendous variation in the rate of tumor development -- providing an understanding of the heterogeneity in tumor sizes and development times that have been observed by epidemiologists and clinicians.   Furthermore, the model provides a simple formula for the number of driver mutations as a function of the total number of mutations in the tumor.   Finally, when applied to recent experimental data, the model allows us to calculate, for the first time, the actual selective advantage provided by typical somatic mutations in human tumors in situ.  This selective advantage is surprisingly small, 0.005 $\pm$ 0.0005,  and has major implications for experimental cancer research.     
\end{abstract}

\maketitle

It is now well-accepted that virtually all cancers result from the accumulated mutations in genes that increase the fitness of a tumor cell over that of the cells that surround it\cite{vogelstein04,greenman07}.  As a result of advances in technology and bioinformatics, it has recently become possible to determine the entire compendium of mutant genes in a tumor\cite{sjoblom06,wood07,parsons08,jones08_science,ley08,mardis09}. Studies to date have revealed a complex genome, with $\sim40 - 80$ amino-acid changing mutations present in a typical solid tumor\cite{sjoblom06,wood07,parsons08,jones08_science}.  For low frequency mutations, it is difficult to distinguish "driver mutations" -- defined as those that confer a selective growth advantage to the cell -- from "passenger mutations"\cite{greenman07,maley04,haber07}.  Passenger mutations are defined as those which do not alter fitness but occurred in a cell that coincidentally or subsequently acquired a driver mutation, and are therefore found in every cell with that driver mutation.   It is believed that only a small fraction of the total mutations in a tumor are driver mutations, but new, quantitative models are clearly needed to help interpret the significance of the mutational data and to put them into the perspective of other lines of cancer research investigation.  

In most previous models of tumor evolution, mutations accumulate in cell populations of constant size\cite{nowak02,nowak04, durrett09} or of variable size, but the models take into account only one or two mutations\cite{iwasa06,haeno07,dewanji05,komarova07,meza08}. In our new model, we assume that each new driver mutation leads to a slightly faster tumor growth rate. This model is as simple as possible as the analytical results depend on only three parameters: the average driver mutation rate, the average selective advantage associated with driver mutations, and the average cell division time. 

Tumors are initiated by the first genetic alteration that provides a relative fitness advantage. In the case of typical leukemias, this would represent the first alteration of an oncogene, such as a translocation between BCR and ABL.  In the case of solid tumors, the mutation that initiated the process might actually be the second "hitÓ in a tumor suppressor gene -- the first hit affects one allele, without causing a growth change, while the second hit, in the opposite allele, leaves the cell without any functional suppressor, in accord with the two-hit hypothesis\cite{knudson71}. It is important to point out that we are modeling tumor progression, not initiation\cite{nowak02,nowak04}, because progression is rate-limiting for cancer mortality -- it generally requires three or more decades for metastatic cancers to develop from initiated cells in humans.   

Our first goal is to characterize the times at which successive driver mutations arise in a tumor of growing size. We have employed a discrete time branching process in this model because it makes the numerical simulations feasible. In a discrete time process, all cell divisions are synchronized. We present analytic formulas for this discrete time branching process and analogous formulas for the continuous time case (in Appendix \ref{cont}) when possible. At each time step, a cell can either divide or differentiate, senesce, or die.  In the context of tumor expansion, there is no difference between differentiation, death, and senescence, as none of these processes will result in a greater number of tumor cells than present prior to that time step.  We assume that driver mutations reduce the probability that the cell will take this second course, i.e., that it will differentiate, die, or senesce, henceforth grouped as   "stagnate".  A cell with $k$ driver mutations therefore  has a stagnation probability $d_k={1\over 2} (1-s)^k$. The division probability is $b_k=1-d_k$. The parameter $s$ characterizes the (average) selective advantage of each driver mutation that occurs following the initiating mutation.

\begin{figure}
\centering
\includegraphics*[width=0.45\textwidth]{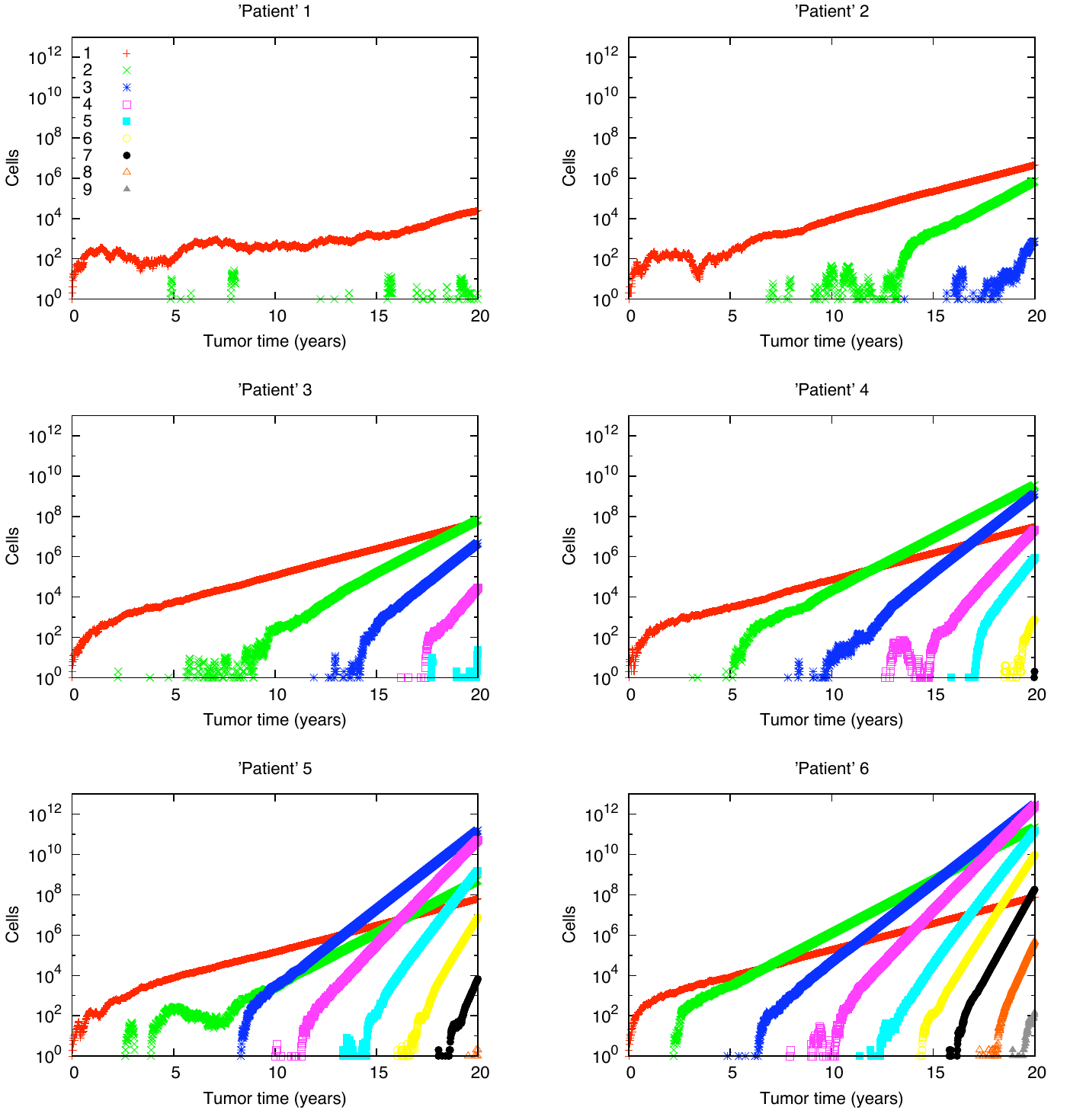}
\caption{ {\bf Variability in tumor progression.} Number of cells with a given number of driver mutations versus the age of the tumor. Six different realizations of the same stochastic process with the same parameter values are shown, corresponding to tumor growth in six 'patients'. The process is initiated with a single surviving founder cell with one driver mutation. The times at which subsequent driver mutations arose varied widely among 'patients'. After initial stochastic fluctuations, each new mutant lineage grew exponentially. The overall dynamics of tumor growth are greatly affected by the random time of the appearance of new mutants with surviving lineages. Parameter values: mutation rate $u = 10^{-5}$, selective advantage $s = 0.5\%$ and generation time $T = 3$ days.}
\end{figure}

When a cell divides, one of the daughter cells can receive an additional driver mutation with probability $u$. The point mutation rate per base per cell division is $\sim 5 \times 10^{-10}$. There is a finite number of tumor suppressor genes and oncogenes that can be mutated at each time point and lead to driver mutations.  We conservatively estimate that there are $\sim100$ tumor suppressor genes and $\sim 100$ oncogenes in a human cell, and that on average each tumor suppressor gene can be inactivated by mutation at $\sim 200$ positions and each oncogene can be activated in $\sim 10$ positions.  There are thus a total of $\sim 21,000$ positions in the genome that could become driver mutations. As the rate of chromosome loss in tumors is much higher than the rate of point mutation\cite{nowak02},  a single point mutation is rate limiting for inactivation of tumor suppressor genes. The driver mutation rate is therefore $\sim 10^{-5}$ per cell division ($\approx 21000 \times 5 \times 10^{-10}$). Our theory can accommodate any realistic mutation rate and the major numerical results are only weakly affected by varying the mutation rate within a reasonable
range.   

Experimental evidence suggests that tumor cells divide about once every three days in glioblastoma multiforme\cite{hoshino79} and once every four days in colorectal cancers\cite{jones08}.  Incorporating these division times into the simulations provided by our model leads to the dramatic results presented in Fig.\ 1.   Though the same parameter values -- $u = 10^{-5}$ and $s = 0.5\%$ -- were used for each simulation, there was an enormous variation in the times required for disease progression.  For example, in 'Patient 1', the second driver mutation had not occurred within 20 years following tumor initiation and the size of the tumor remained small (micrograms, representing   $< 10^5$ cells). In contrast, in 'Patient 6', the second driver mutation occurred after only two years and by 20 years the tumor would weigh kilograms ($10^{12}$ cells), with the most common cell types in the tumor having three or four driver mutations.   'Patients 2 to 5' had progression rates between these two extreme cases.

The results in Fig.\ 1 provide insights into the nature of tumor development in patients with familial adenomatous polyposis (FAP)\cite{muto75}.  If untreated (by colectomy), these patients develop adenomas while teenagers, but do not develop cancers until their fourth or fifth decades of life, by which time there are thousands of tumors per patient. Each of these thousands of tumors represents a process that is roughly similar to one of the processes in Fig.\ 1.  Hence it is likely that at least one of them will progress relatively fast, as in 'Patient 6' in Fig.\ 1.   A patient without FAP who has initiated one polyp has only a $\sim 0.001$ probability of having a cell in that polyp that has accumulated ten driver mutations, even after 20 years. On the other hand, a patient that has initiated 1000 polyps has a 0.63 probability of having a cell in at least one polyp with ten driver mutations during the same time period.  This quantitatively explains (i) why the polyps in an individual patient with FAP are so heterogeneous in size and (ii) why cancers are so much more common in FAP patients than in the general population, despite the fact that the inherited mutation affects only tumor initiation and does not directly affect tumor progression. 

We can calculate the average time between the appearance of successful cell lineages (Fig.\ 2). Not all new mutants are successful, because stochastic fluctuations can lead to the extinction of a lineage. The lineage of a cell with $k$ driver mutations  survives only with a probability approximately $1-d_k/b_k\approx 2sk$. Assuming that $u\ll ks\ll 1$, the average time between the first successful cell with $k$ and the first successful cell with $k+1$ driver mutations is given by
$$\tau_k = {{T}\over{ks}} \log {{2ks}\over u}. \eqno(1)$$
The introduction of subsequent driver mutations becomes faster and faster. (See Appendices \ref{rate} and \ref{waiting})
For example, for $u=10^{-5}$,  $s=10^{-2}$ and $T=4$ days it takes on average  8.3 years until the 2nd driver mutation emerges, but only 4.5 more years until the 3rd driver mutation emerges. The cumulative time to have $k$ mutations grows logarithmically with $k$. See Table 1 for additional numerical examples. 

In contrast to driver mutations, passenger mutations do not confer a fitness advantage, and they do not modify tumor growth rates. We find that the average number of passenger mutations, $n(t)$, present in a tumor cell after $t$ days is proportional to $t$, that is $n(t) = v t/T$, where $v$ is the rate of acquisition of neutral mutations. In fact, $v$ is the product of the point mutation rate per base pair and the number of base pairs analyzed. This simple relation has been used to analyze experimental results by providing estimates for relevant time scales\cite{jones08}.  

Combining our results for driver and passenger mutations, we can derive a formula for the number of passengers that are expected in a tumor that has accumulated $k$ driver mutations (see Appendix \ref{comb})
$$n = \frac{v}{2s} \log{\frac{4ks^2}{u^2}}\log k. \eqno(2)$$
Here $n$ is the number of passengers that were present in the last cell that clonally expanded.  Equation (2) can be most easily applied to tumors in tissues in which there was not much cell division prior to the tumor initiation, as otherwise a difficult estimate of the expected number of passengers that accumulated in a precursor cell prior to tumor initiation would be required.

 \begin{figure}
\centering
\includegraphics*[width=0.45\textwidth]{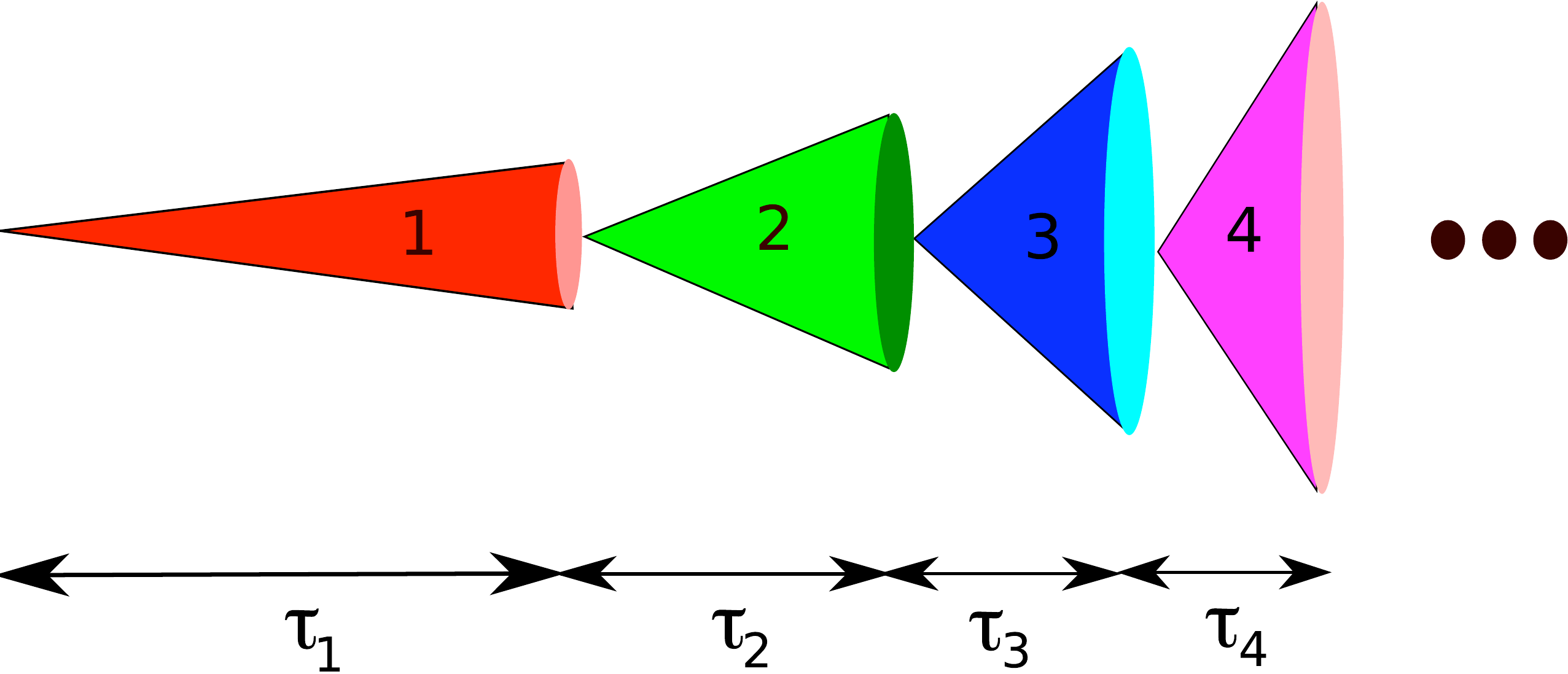}
 \caption{ {\bf Schematic representation of waves of clonal expansions.}
An illustration of a sequence of clonal expansions of cells with $k=$1, 2, 3 or 4 driver mutations is shown. Here $\tau_1$ is the average time it takes the lineage of the founder cell to produce a successful cell with two driver mutations.  Similarly, $\tau_k$ is the average time between the appearance of cells with $k$ and $k+1$ mutations. Equation (1) gives a simple formula for these waiting times, which shows that subsequent driver mutations appear faster and faster. The cumulative time to have $k$ driver mutations grows with the logarithm of $k$.}
\end{figure}

\begin{table}
\begin{center}
\begin{tabular}{|cc|cccc|}
	\hline 
$s$  & $u$ &  $\tau_1$    &   $\tau_2$  & $\tau_3$ & $\tau_4$    \\
	\hline 
$0.1\%$      & $10^{-5}$ & $58.0$  & $32.8$ & $23.4$ &$18.3$ \\

$0.5\%$   & $10^{-5}$ & $15.1$ & $8.3$ & $5.8$ & $4.5$ \\

$1\%$   & $10^{-5}$ & $8.3$ & $4.5$ & $3.2$  & $2.5$ \\

$2\%$   & $10^{-5}$ & $4.5$ & $2.5$ & $1.7$  & $1.3$ \\

$10\%$   &$10^{-5}$ & $1.1$ & $0.6$ & $0.4$ & $0.3$ \\

	\hline
$1\%$      & $10^{-6}$ & $10.8$ & $5.8$ & $4.0$ & $3.1$ \\

$1\%$   & $5 \cdot 10^{-6}$ & $9.1$ & $4.9$ & $3.4$ & $2.7$\\

$1\%$   & $10^{-5}$ & $8.3$ & $4.5$ & $3.2$  & $2.5$ \\

$1\%$   & $2 \cdot 10^{-5}$ & $7.6$ & $4.2$ & $2.9$ & $2.3$\\

$1\%$   &$10^{-4}$ & $5.8$ & $3.3$ & $2.3$ & $1.8$ \\

	\hline
\end{tabular}
\end{center}
\caption{{\bf Times between clonal waves}
Numerical values for the average time $\tau_k$ (in years) between the first successful cell with $k$ and $k+1$ driver mutations, for different values of the selective advantage $s$ and the mutation rate $u$. Cells divide every $T=4$ days. The table shows that changing the selective advantage of drivers has a large effect on the waiting times, while changing the driver mutation rate has a relatively small effect.}
\end{table}

\begin{figure}
\centering
\includegraphics*[width=0.5\textwidth]{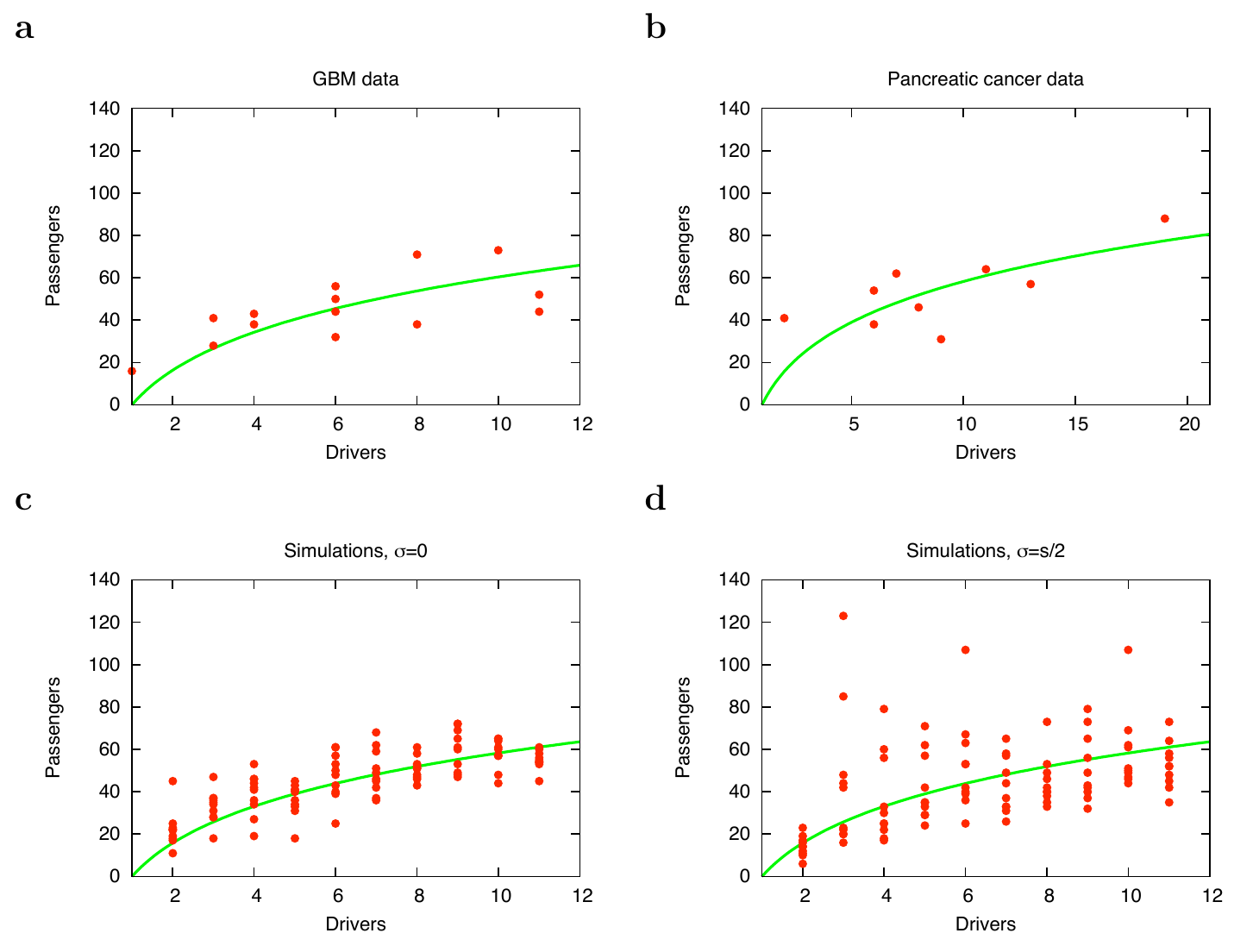}
\caption{ {\bf Comparison of clinical mutation data and theory.}
Our theory provides an estimate for how the number of passenger mutations found in a tumor is related to the number of driver mutations. Here we show a comparison of equation (2) (green line) with genomic data from glioblastoma multiforme (GBM) and pancreatic cancer. We also compare our analytic result with computer simulations. Parameter values used in equation (2) and computer simulations were $s = 0.5\%$ and $u = 10^{-5}$. {\bf a}, Equation (2) (green line) fitted to GBM data from 14 patients. {\bf b}, Equation (2) (green line) fitted to pancreatic cancer data from 9 patients. {\bf c}, Comparison of computer simulations and equation (2). For each $k$ between 2 and 10,  the number of passengers that were brought along with the last driver in 10 tumors with $k$ drivers is plotted. {\bf d}, Comparison between computer simulations and equation (2) for selective advantage of the $k$-th driver, $s_k$, taken from a Gaussian distribution with mean $s$ and standard deviation $\sigma = s/2$. For each $k$ between 2 and 10, the number of passengers that were brought along with the last driver in 10 tumors with $k$ drivers is plotted.}
\end{figure}

To test the validity of this model, we tested it on two tumor types that have been extensively analyzed.    Neither the astrocytic precursor cells that give rise to glioblastoma multiforme (GBM)\cite{louis07} nor the pancreatic duct epithelial cells that give rise to pancreatic adenocarcinomas\cite{mimeault05} divide much prior to tumor initiation\cite{kraus-ruppert73,klein02}.  Therefore the data on both tumors should be suitable for our analysis.     Parsons {\it et al.}\cite{parsons08} sequenced 20,661 protein coding genes in a series of GBM tumors and found a total of 713 somatic mutations in the 14 samples that are depicted in Fig.\ 3.  Similarly, Jones {\it et al.}\cite{jones08_science} sequenced the same genes in a series of pancreatic adenocarcinomas, finding a total of 562 somatic mutations in the 9 primary tumors graphed in Fig.\ 3. In both cases, we classified missense mutations as drivers if they were classified as such by the CHASM algorithm at false discovery rate (FDR) 0.2\cite{carter09} (see Appendices \ref{data} and \ref{chasm}).  We also considered all nonsense mutations, out-of-frame insertions or deletions (INDELs) and splice-site changes as drivers because these generally lead to inactivation of the protein products\cite{jones08_science}.  All other somatic mutations were considered to be passengers.  

From Fig.\ 3a and b, it is clear that the experimental results on both GBM and pancreatic cancers were in good accord with the predictions of equation (2).  A critical test of the model can be performed by comparison of the best-fit parameters governing each tumor type.  It is expected that the average selective advantage of a driver mutation should be similar across all tumor types given that the pathways through which these mutations act overlap to a considerable degree.   Setting the driver mutation rate to be $u = 10^{-5}$ and fitting equation (2) to the GBM data using least squares analysis, we found that the optimum fit was given by $s = 0.0048 \pm 0.0004$.   Remarkably, using the same mutation rate in pancreatic cancers, we find that the best fit is given by a nearly identical $s=0.0050 \pm 0.0005$. This consistency not only provides support for the model but also provides evidence that the average selective advantage of a driver is s $\approx 0.5\%$. For $u = 10^{-6}$ and $u = 10^{-4}$, we get $s\approx0.65\%$ and $s\approx0.32\%$, respectively.   The fact that these estimates are not strongly dependent on the mutation rate supports the robustness of the model.

In Beerenwinkel {\it et al.}\cite{beerenwinkel07}, we have  previously modeled tumors which are slowly expanding due to some constraint, using a Wright-Fisher process. The new results are considerably different.  For example, Beerenwinkel {\it et al.} found that the "waiting timeÓ required to accumulate $k$ driver mutations was proportional to $k$, while in the new model the waiting time depends on the logarithm of $k$.  Our formula for the waiting time provides a much better explanation for the long initial stages in the adenoma-carcinoma sequence\cite{jones08}. 

Like all models, ours incorporates limiting assumptions. However, many of these assumptions can be loosened without changing the key conclusions. For example, we assumed that the selective advantage of every driver was the same. We have tested whether our formulas still hold in a setting wherein the selective advantage of the $k$-th driver is $s_k$, and $s_k$'s are drawn from a Gaussian distribution with mean $s$ and standard deviation $\sigma = s/2$.  The simulations were still in excellent agreement with equation (2) (Fig.\ 3d).  Similarly, we assumed that the time between cell divisions (generation time T) was constant. Nevertheless, equation (2), which gives the relationship between drivers and passengers, is derived without any specification of time between cell divisions. Consequently, this formula is not affected by the possible change in T.  Finally, there could be a finite carrying capacity for each mutant lineage. In other words, cells with one driver mutation may only grow up to a certain size, and the tumor may only grow further if it accumulates an extra mutation, allowing cells with two mutations to grow until they reach their carrying capacity and so on. It is reasonable to assume that the carrying capacities of each class would be much larger than $1/u$, which is approximately the number of cells with $k$ mutations needed to produce a cell with $k + 1$ mutation.  Thus, the times at which new mutations arise would not be much affected by this potential confounding factor.

Given the true complexity of cancer, our model is deliberately oversimplified. Despite its simplicity, however, it captures some essential characteristics of the genetic complexity underlying tumor growth. Simple models have already been very successful in providing important insights into cancer. Notable examples include Knudson's two hit model\cite{knudson71} and Armitage-Doll's multi-hit hypothesis\cite{armitage04}. The model described here represents the first attempt to provide analytical insights into the relationship between drivers and passengers in tumor progression and will hopefully be similarly stimulating.  One of the major conclusions, i.e., that the selective growth advantage afforded by the mutations that drive tumor progression is very small ($\sim 0.5\%$), has major implications for understanding tumor evolution.  For example, it shows how difficult it will be to create valid in vitro models to test such mutations on tumor growth; selective growth advantages of $0.5\%$ are nearly impossible to discern in cell culture over short time periods.  And it explains why so many driver mutations are needed to form an advanced malignancy within the lifetime of an individual.

\subsection*{Acknowledgements} This work is supported by the John Templeton Foundation, the NSF/NIH (R01GM078986) joint program in mathematical biology, the Bill and Melinda Gates Foundation (Grand Challenges Grant 37874), and J. Epstein.

\appendix

\section{Simulations}
\label{simu}

We model tumor progression with a discrete time Galton-Watson branching process.
In our model, at each time step a cell with $j$ mutations (or $j$-cell) either divides into two cells, which occurs with probability $b_j$, or dies with probability $d_j$, where $b_j+d_j=1$. In addition, at every division, one of the daughter cells can acquire an additional mutation with probability $u$. The process is initiated by a single cell with one mutation. We set $d_j=\frac{1}{2}(1-s)^j$, so that additional mutations reduce the probability of cell death. 

In simulations, we track the numbers of cells with $j$ mutations, $N_j$, for $j=1,\dots,p$, rather than the faith of each individual cell. We increase the efficiency of the computation by sampling from the multinomial distribution at each time step. Let $N_j(t)$ be the number of cells with $j$ mutations at time $t$. Then the number of $j$-cells that will give birth to an identical daughter cell, $B_j$, the number of $j$-cells that will die, $D_j$, and the number of $j$-cells that will give birth to a cell with an extra mutation, $M_j$, are sampled from the multinomial distribution with
\begin{equation}
\begin{split}
\mathrm{Prob}&[(B_j,D_j,M_j)=(n_1,n_2,n_3)] = \\
 &=  \frac{N_j(t)!}{n_1!n_2!n_3!} [b_j(1-u)]^{n_1} d_j^{n_2} (b_ju)^{n_3},
\end{split}
\end{equation}
for $n_1+n_2+n_3=N_j(t)$. Then, 
\begin{equation}
N_j(t+1)=N_j(t)+B_j-D_j+M_{j-1}.
\end{equation}
Note that in this model, all cell divisions and cell deaths occur simultaneously at each time step. One could define an analogous continuous time model, with a very similar behavior. Simulations of the continuous time model, however, are much less efficient, since the updates occur at smaller and smaller time steps as the population size grows.

\section{Average number of $j$-cells}

In this section, we derive an analytical expression for the average abundance of  cells with $j$ mutations after $t$ time steps in a Galton-Watson process. Since the population is initiated by a single cell, it might go extinct due to random events. If the population survives, it exhibits waves of clonal expansions.

Let $x_j(n)$ denote the average number of $j$-cells after $n$ time steps. The averages $x_j$ are evolving according to the following system of difference equations:
\begin{equation}
\label{avereqs}
\begin{split}
x_1(n+1)&=b_1(2-u)x_1(n)\\ 
x_j(n+1)&=b_j(2-u)x_j(n)+b_{j-1}ux_{j-1}(n), \,\,\,\,\,\,\,\,  j>1,
\end{split}
\end{equation} 
with initial conditions $x_1(0)=1$ and $x_j(0)=0$ for $j>1$.
This system of infinite number of linear differential equations can be solved exactly. For a single initial 1-cell, we obtain
\begin{equation}
x_1(n)=[b_1(2-u)]^n.
\end{equation}
From this, we can find the solution for 2-cells and so on, arriving at the formula
for the expected number of $j$-cells after $n$ time steps 
\begin{equation}
\label{ex_formula}
x_j(n)=u^{j-1}(2-u)^{n-j+1}  \prod_{k=1}^{j-1} b_k \sum_{k=1}^j \frac{b_k^n}{\prod_{q=1,q \neq k}^j b_k-b_q}.
\end{equation}
The expression \eqref{ex_formula} is the exact formula for the average number of cells with $j$ mutations in our process, which holds for any choice of selective advantage $s$ and mutation rate $u$ between 0 and 1. This can be easily proven by substituting the solution \eqref{ex_formula} into the equations \eqref{avereqs}.
This formula can be approximated by a simpler analytical expression in the small $s$ and  $u \ll s$ limit, which are reasonable assumptions for tumor progression.

Since $b_k$'s are strictly increasing with $k$, $x_j(t)$ can be well approximated by its leading term behavior,
\begin{equation}
\label{leading_term}
x_j(t)=a_j [b_j(2-u)]^t ,
\end{equation}
with $$a_j=\left(\frac{u}{2-u}\right)^{j-1} \left(\prod_{k=1}^{j-1} \frac{b_k}{b_j-b_k}\right),$$
for all  times $t$ for which $e^{-st} \ll 1$.
In the weak selection limit, $s \ll 1/j$ and in the low mutation limit $u\ll 1$, we can approximate $(1-s)^j \approx 1-js$ and $(2-u) \approx 2$, to get
\begin{equation}
\label{ap_formula_aj}
a_j = \frac{1}{(j-1)!}\left(\frac{u}{2s}\right)^{j-1}.
\end{equation}
In addition, if $u \ll s$ we have $[b_j(2-u)]^t \approx [2-(1-s)^j]^t$.
leading to the expression for the average abundance of $j$-cells after $t$ time steps
\begin{equation}
\label{ap_formula}
x_j(t) = \frac{1}{(j-1)!}\left(\frac{u}{2s}\right)^{j-1}\left[2-(1-s)^j\right]^t.
\end{equation}
Note that in the last equation we do not approximate $(1-s)^j$ with $1-js$, since for large values of $t$ this approximation would lead to significant errors. 

 \begin{figure}
\centering
\includegraphics[scale=0.85]{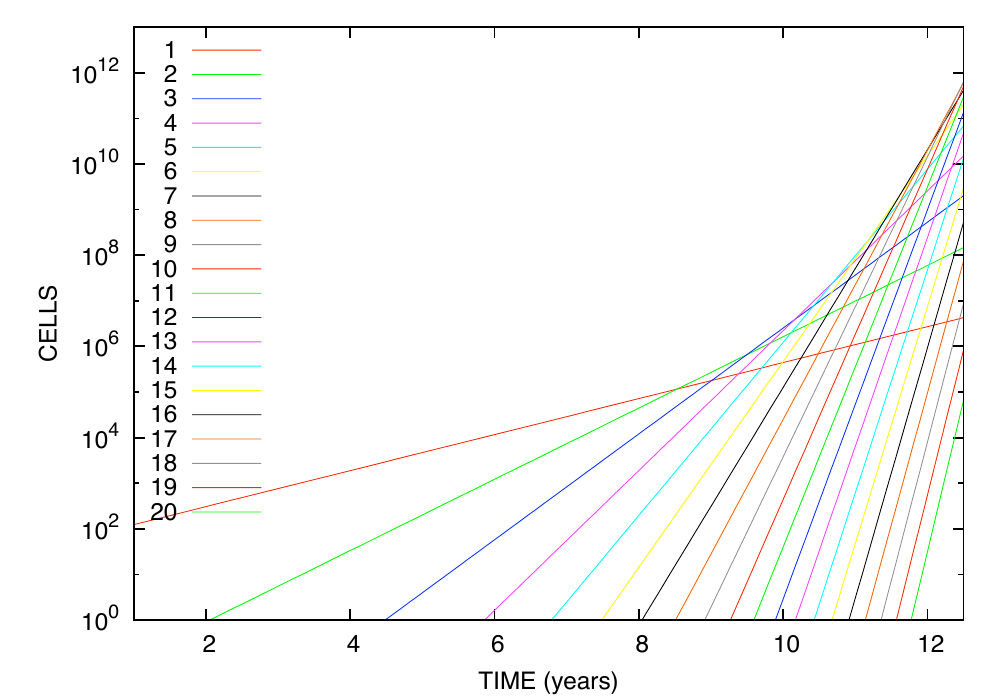}
\label{fig:averages}
\caption{Average number of cells with $j$ mutations, as predicted by formula \eqref{ap_formula}.  Parameter values are $s=0.01$, $u=10^{-5}$ and $T=4$ days.}
\end{figure}

Formulas \eqref{ex_formula} and \eqref{ap_formula} are averages calculated considering both trajectories of extinction and non-extinction. The average abundance of cells with $j$ mutations after $t$ time steps, conditioned on non-extinction, is given by
\begin{equation}
X_j(t)  =  \frac { x_j(t)}{1-q(t)},
\end{equation}
where $q(t)$ is the is the probability that the population goes extinct by $t$. We note that extinction is most likely an early event in the process and that it occurs when the population consists almost entirely of 1-cells. Thus we can approximate $q(t)$ by eventual extinction probability of a population initiated by a single cell with one mutation, in a process with no mutation  
\begin{equation}
\label{ext_ap}
q(t) \approx q_1.
\end{equation}
In a process with no mutation, the probability that a population initiated by a single cell with one mutation goes extinct is given by
\begin{equation}
\label{extinction}
q_1=\frac{d_1}{1-d_1}=\frac{1-s}{1+s} \approx 1-2s
\end{equation}
where the last approximation is valid in the small $s$ limit.

Finally we have the expression for the average number of cells with $j$ mutations, conditioned on non-extinction,
\begin{equation}
\label{ap_formula_nex}
X_j(t) = \frac{1}{2s(j-1)!}\left(\frac{u}{2s}\right)^{j-1}(2-(1-s)^j)^t.
\end{equation}

The average numbers of cells with $j$ mutations for $j=1, \dots, 20$, as predicted by formula \eqref{ap_formula_nex}, are shown in Fig.\ S1. In Fig.\ S2, we plot the excellent agreement between these predicted averages and simulations. However, we note that due to large fluctuations in the numbers of cells with $j$ mutations at time $t$ (see Fig.\ 1 from the main text), the averages \eqref{ap_formula_nex} are not that useful for describing the dynamics of the process, as they are biased towards realizations with large numbers of cells. For example, in Fig.\ S1 we see that the average number of cells with 2 mutations is 1 after about 2 years. On the other hand, simulation suggest that the average time to the first successful cell with 2 mutations is approximately 8 years for the parameter values from Fig.\ S1.

 \begin{figure}
\centering
\includegraphics[scale=1.2]{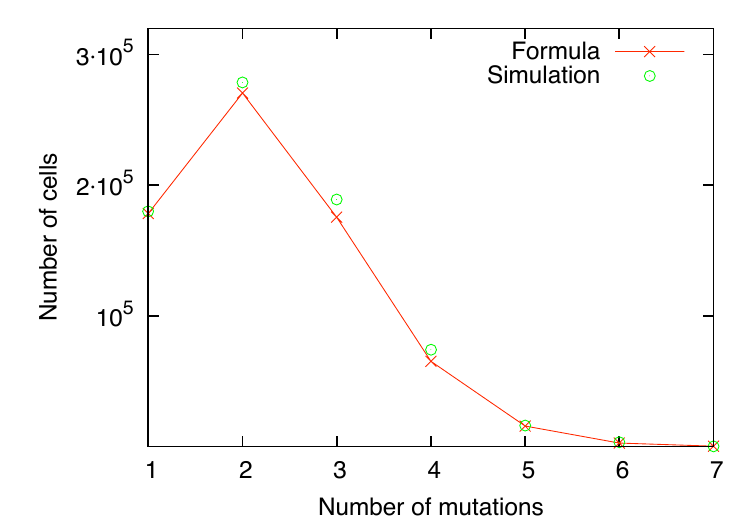}
\label{fig:averages_formula_sim}
\caption{Average number of cells with $j$ mutations, comparison of formula \eqref{ap_formula} and simulations.  Parameter values are $s=0.01$, $u=10^{-5}$, $T=4$ days and $t=9$ years.}
\end{figure}

\section{The rate of introduction of new mutants}
\label{rate}

Simulations of our Galton-Watson process suggest that the times at which a new mutant with a surviving lineage is produced have a significant effect on the dynamics of the process. In this section we give an approximation for the average time it takes the first $j$-cell with surviving lineage to produce a $(j+1)$-cell with surviving lineage. 

The average number of $j$-cells grows as $x_j=\frac{1}{1-q_j} [b_j (2-u)]^{\tau}$, where $\tau$ is the time measured from the appearance of the first successful $j$-cell and $q_j=\frac{d_j}{1-d_j}$ is the extinction probability of $j$-cells. New $(j+1)$-cells with surviving lineages appear at rate $(1-q_{j+1})ub_jx_j$, and we approximate the time of the appearance of the first $(j+1)$ cell with  surviving lineage, $\tau_j$, by the time when the total rate reaches one cell, that is
\begin{equation}
 \sum_{k=1}^{\tau_j}\frac{1-q_{j+1}}{1-q_j} u b_j [b_j(2-u)]^k = 1
\end{equation}
This leads to 
\begin{equation}
 \tau_j = \frac{\log \left[1+ \frac{1-q_j}{ub_j(1-q_{j+1})}\left(1- \frac{1}{b_j(2-u)}\right)\right]} {\log [b_j(2-u)]}.
\end{equation}

\begin{figure}
\centering
\includegraphics[scale=1.2]{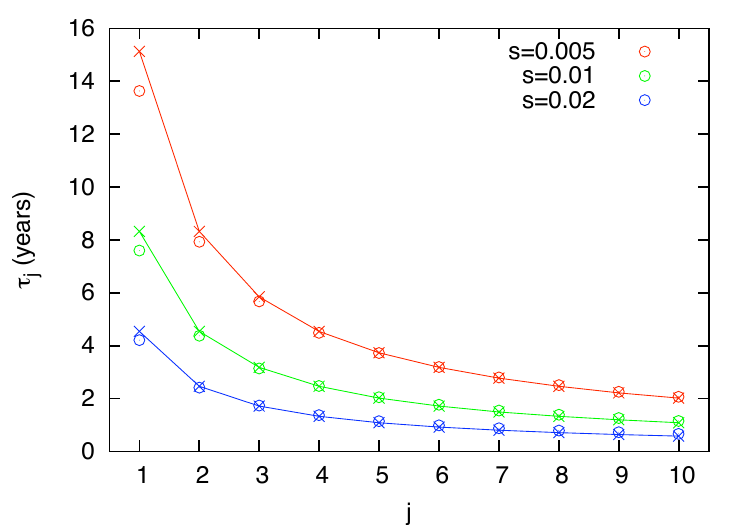}
\label{fig:tau_j}
\caption{Speed of introduction of new mutants: comparison of formula and simulation.  Comparison of predicted and simulated average time it takes the lineage of the first successful  $j$-mutant to produce the first successful $(j+1)$-mutant, $\tau_j$, for different values of selective advantage $s$. Circles correspond to times obtained from simulations, and lines correspond to formula \eqref{tau_j_ap}. Parameter values are $u=10^{-5}$ and $T=4$ days.}
\end{figure}

We consider selection and mutation rate to be small enough, $u\ll 1$ and $s\ll 1$, so $\log [b_j(2-u)] \approx js $. We also assume $js\ll 1$ so we can approximate  $(2-(1-s)^j) \approx 1+js$, and thus $1- 1/[b_j(2-u)] \approx js$ . In these limits we also have $\frac{1-q_j}{ub_j(1-q_{j+1})} \approx \frac{2j}{u(j+1)}$. Now we can write
\begin{equation}
\label{tau_j}
 \tau_j =\frac {\log \frac{2j^2 s}{(j+1)u}}{js}.
\end{equation}
We can further simplify this formula by noting that $\frac{j}{j+1} \approx 1$ to obtain
 \begin{equation}
\label{tau_j_ap}
 \tau_j =\frac {\log \frac{2j s}{u}}{js}.
\end{equation}
If, in addition, we assume that generation time is $T$, we get
 \begin{equation}
\label{tau_j_ap_realtime}
 \tau_j =\frac{T}{js} \log \frac{2j s}{u}.
\end{equation}

The excellent agreement between formula \eqref{tau_j_ap} and simulations is shown if Fig.\ S3.

\section{Waiting time to $k$ mutations}
\label{waiting}

We also derive a formula for the average time it takes for the first successful $k$-mutant to be produced in the process, $t_k$, by assuming
 \begin{equation}
 t_k=\sum_{j=1}^{k-1} \tau_j.
 \end{equation}
Substituting expression \eqref{tau_j_ap} for $\tau_j$, we arrive at
  \begin{equation}
 t_k=\sum_{j=1}^{k-1} \frac {\log \frac{2j s}{u}}{js}.
 \end{equation}
 We approximate the last sum with an integral as
  \begin{equation}
 t_k=\int_1^k \frac {\log \frac{2j s}{u}}{js}.
 \end{equation}
which then leads to the following formula for waiting time
\begin{equation}
\label{timecum}
 t_k =\frac {T}{2s} \, \log \frac{4ks^2}{u^2} \, \log k.
\end{equation}
The comparison between formula \eqref{timecum} and simulations is shown if Fig.\ S4. 

\begin{figure}
\centering
\includegraphics[scale=1.2]{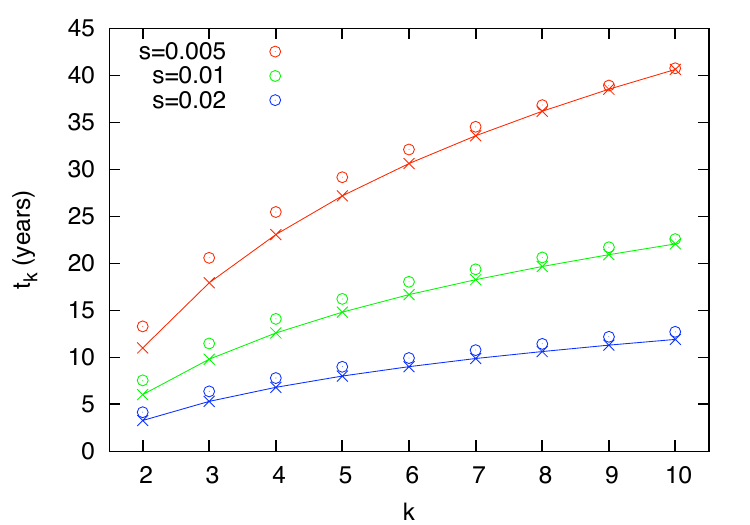}
\label{fig:t_k}
\caption{Waiting time to $k$ mutations. Comparison of predicted and simulated average time it takes for the first successful  $k$-mutant to be produced in the process for different values of selective advantage $s$. Circles correspond to times obtained from simulations, and lines correspond to formula \eqref{timecum}. Parameter values are $u=10^{-5}$ and $T=4$ days.}
\end{figure}
 
\section{Passenger mutations}
\label{pass}

Suppose now that we have a model in which there are two types of mutations: drivers, which confer selective advantage as before, and passengers, which have no influence on the fitness of the cell. If a cell with $n$ passenger mutations divides, then each of the daughter cells can have one additional passenger mutations with probability $v$. Since passenger mutation do not affect the fitness of the cell, after $t$ time steps, each cell still alive has the probability 
\begin{equation}
\label{neutdis}
{t \choose n} v^n (1-v)^{t-n}
\end{equation}
to have $n$ passenger mutations. It follows that the average number of passenger mutations present in the neoplastic cell population after $t$ time steps is 
 \begin{equation}
 \label{pass}
 n(t)=t v.
\end{equation}
Note that a crucial condition for \eqref{neutdis} to be valid is that the time increments must be constant, that is by time $t$ each cell undergoes $t$ cell divisions. This condition is not satisfied generally in continuous time branching processes. Note also that, while in our model only one of the two offsprings can acquire a driver mutation in a cell division, both of them can acquire a passenger mutation. The reason is that we safely neglected the possibility of new driver mutations in both offsprings, since that is roughly $u/2=0.5\times10^{-5}$ times less probable than acquiring a driver mutations in only one of the offsprings.

\section{Drivers vs passengers}
\label{comb}

Combining our results \eqref{timecum} and \eqref{pass} for driver and passenger mutations, we give a formula for the number of passengers we expect to find in a tumor that accumulated $k$ driver mutations
\begin{equation}
\label{driv_pass}
n = \frac{v}{2s} \log{\frac{4ks^2}{u^2}}\log k.
\end{equation}

Note that $n$ is the number of passengers that were present in the last cell that clonally expanded. It is these passenger mutations that can be detected experimentally. Formula \eqref{driv_pass} can only be applied to tumors in tissues in which there was not much cell division prior to tumorigenesis.

\section{Continuous time formulas}
\label{cont}

In this section we define a similar continuous time model and list the above analytical results in this setting. As before, we start with one cell with one driver mutation. In a short time interval $\Delta t$, a cell with $j$ driver mutations can divide with probability $b_j \Delta t$ and die with probability $d_j \Delta t$. 

Let $x_j(t)$ be the expected number of cells with $j$ mutations alive at time $t$. The population is evolving according to the differential equations:
$$\frac{dx_1}{dt}=[b_1(1-u)-d_1]x_1,$$
$$\frac{dx_j}{dt}=[b_j(1-u)-d_j]x_j + b_{j-1}u x_{j-1}   \ \  \rm{for} \ \ \emph{j} \geq 2,$$
subject to the initial conditions  $x_1(0)=1$ and $ x_j(0)=0$ for $j \geq 2$.
This system can be solved analytically and the exact solution is
\begin{equation}
\label{solspec}
 x_j = \prod_{l=1}^{j-1} b_l u
 \sum_{k=1}^{j}  \frac{e^{r_k t}}{\prod_{q=1, q\ne k}^{j} ((b_k-b_q)(1-u)+d_q - d_k) }
\end{equation}
with $r_k=b_k(1-u)-d_k$.

In order to model tumor progression, let us specify the rates $b_j$ and $d_j$. Perhaps the simplest choice is to assign the same fitness advantage to each driver mutations, that is have a $j$ dependent division rate $b_j=1+sj$, and constant death rate $d_j=1$. The main problem with this choice is, that when substituting them into the general formula \eqref{solspec}, it turns out that the average number of cells becomes infinite at the finite time $t^*=-\log u/[s(1-u)]$. The underlying reason for this blowup is the presence of an infinite number of cell types. This artifact can be easily avoided by making each mutation decrease the death rate of cells, that is to define $d_j=(1-s)^j$, and to make the division rate constant $b_j=1$. The population always remains finite in this version of the model. Fitter cells, however, have shorter generation times than less fit cells. Hence, at any given time $t$, different cells may have undergone different numbers of cell divisions.  As a consequence, the expected number of neutral mutations is not the same for all cells (in fact it is positively correlated with the number of driver mutations), hence we do not have a simple relationship between drivers and passengers as in the discrete time case. For this reason we propose the following definition instead.

We define a continuous time branching process similar to the discrete one we use in the paper. In this process, an event (division or death of a cell) occurs at rate $1/T$. If an event occurs to a cell with $j$ mutations, then it is death with probability $\frac{1}{2} (1-s)^j$ and division with probability $1-\frac{1}{2} (1-s)^j$. Thus, $b_j=(1-\frac{1}{2} (1-s)^j)/T$ and $d_j=\frac{1}{2T} (1-s)^j$.

In this case,  the time between the appearance of the first successful $j$-cell and the appearance of the first successful $(j+1)$ cell, $\tau_j$ is given by
 \begin{equation}
\label{tau_j_ap_c}
 \tau_j =\frac{T}{js} \log \frac{2 j s}{u T}.
\end{equation}
The waiting time to the first successful $k$ mutation is 
\begin{equation}
\label{timecum_c}
 t_k =\frac {T}{2s} \, \log \frac{4ks^2}{(uT)^2} \, \log k.
\end{equation}

Since the times between successive divisions of a single cell line are constant on average, we can use formula \eqref{pass} for passenger mutations, in order to get the following formula for the number of passengers as a function of the number of drivers
\begin{equation}
\label{pass_c}
 n =\frac {v}{2s} \, \log \frac{4ks^2}{(uT)^2} \, \log k.
\end{equation}

\section{Mutation data}
\label{data}

\begin{table}
\begin{center}
\begin{tabular}{||cccc||}
	\hline \hline
Gene  & Mutation &  CHASM score    &  $P$-value     \\
	\hline \hline
CDKN2A	& H98P	& 0.024 &	0.0004 \\
CDKN2A	& L63V	& 0.096 &	0.0004 \\
TP53 &	C275Y &	0.028 &	0.0004 \\
TP53 &	G266V &	0.024 &	0.0004 \\
TP53 &	H179R &	0.152 &	0.0004 \\
TP53 &	I255N &	0.024 &	0.0004 \\
TP53 &	L257P &	0.048 &	0.0004 \\
TP53* &	R175H &	0.078 &	0.0004 \\
TP53* &	R248W &	0.114 &	0.0004 \\
TP53 &	R282W &	0.126 &	0.0004 \\
TP53 &	S241F &	0.044 &	0.0004 \\
TP53* &	V217G &	0.144 &	0.0004 \\
TP53* &	Y234C &	0.022 &	0.0004 \\
NEK8 &	A197P &	0.268 &	0.0008 \\
PIK3CG &	R839C &	0.258 &	0.0008 \\
SMAD4* &	C363R &	0.240 &	0.0008 \\
TP53 &	D208V &	0.240 &	0.0008 \\
TP53* &	K120R &	0.262 &	0.0008 \\
TP53 &	T155P &	0.202 &	0.0008 \\
MAPT &	G333V &	0.322 &	0.0021 \\
DGKA &	V379I &	0.336 &	0.0025 \\
STK33 &	F323L &	0.342 &	0.0025 \\
FLJ25006 &	S196L &	0.392 &	0.0038 \\
PRDM5* &	V85I	 &  	0.396 &	0.0038 \\
TP53 &	L344P &	0.406 &	0.0050 \\
TTK	&	D697Y &	0.426 &	0.0063 \\
NFATC3* &	G451R &	0.464 &	0.0067 \\
PRKCG* &	P524R &	0.444 &	0.0067 \\
CMAS &	I275R &	0.474 &	0.0071 \\
KRAS* &	G12D &	0.474 &	0.0071 \\
PCDHB2 &	A323V &	0.476 &	0.0071 \\
STN2 &	I590S &	0.474 &	0.0071 \\
SMAD4 &	Y95S &	0.496 &	0.0092 \\
\hline \hline
\end{tabular}
\end{center}
\caption{Driver mutations predicted by CHASM.
Missense mutations found in 24 pancreatic cancer samples from Jones et al.\cite{jones08_science}  which are classified as drivers by CHASM at FDR of 0.2, shown with their associated Random Forest scores and $P$ values. 
(* denotes the missense mutations classified as drivers in the 9 samples used in our analysis.)}
\end{table}

Parsons et al. \cite{parsons08} sequenced 20,661 protein coding genes in 22 human glioblastoma multiforme GBM tumor samples using polymerase chain reaction (PCR) sequence analysis. 7 samples were extracted directly from patient tumors and 15 samples were passaged in nude mice as xenografts. All samples were matched with normal tissue from the same patient in order to exclude germline mutations. Analysis of the identified somatic mutations revealed that one tumor (Br27P), form a patient previously treated with radiation therapy and temozolomide, had 17 times as many alterations as any of the other 21 patients, consistent with previous observations of a hypermutation phenotype in glioma samples of patients treated with temozolomide \cite{cahill07}. After removing Br27P from consideration, it was found that the 6 DNA samples extracted directly from patient tumors had smaller numbers of mutations than those obtained from xenografts, likely because of the masking effect of nonneoplastic cells in the former \cite{jones08}. For this reason we chose only to focus on the mutation data which were taken from xenografts. From the 15 xenograft samples, we excluded one sample(Br04X) because it was taken from a recurrent GBM which may have had prior radiation therapy or chemotherapy, leaving us with 14 samples we used for our study. 

Similarly, Jones et al. \cite{jones08_science} sequenced 20,661 protein coding genes in 24 pancreatic cancers. 10 samples were passaged in nude mice as xenografts and 14 in cell lines.  For the purpose of  our study, we discarded the samples taken from metastases, and used the 9 samples which were taken from primary tumors as xenografts, for consistency with GBM data.

\section{CHASM analysis of missense mutations found in pancreatic cancers}
\label{chasm}

Carter et al. \cite{carter09} used CHASM algorithm to analyse GBM missense mutations found in 22 GBM samples from Parsons et al \cite{parsons08} and classify them as either drivers or passengers. We carried out CHASM analysis of missense mutations found in the original 24 pancreatic cancer samples \cite{jones08_science}.
33 mutations that were classified as drivers by the CHASM algorithm at false discovery rate (FDR) 0.2 are shown in Table S1.

\end{document}